\documentclass{ws-ijmpc}
\usepackage{color}
\usepackage{epstopdf}

\def \etal {et~al.\,}

\def \timStep {t}
\def \posX {x}
\def \posY {y}

\def \posH {h}

\def \velX {u}
\def \velY {v}

\def \dt {\partial_t}
\def \dx {\partial_x}
\def \dxx {\partial_{xx}}
\def \dy {\partial_y}

\def \feq {f^{\mathrm{eq}}}
\def \ueq {{\textbf{u}}^{\mathrm{eq}}}
\def \ci {\vect{c}_i}
\def \cs {c_\mathrm{s}}
\def \scG {G}

\def \ciy {c_{iy}}
\def \cix {c_{ix}}

\def \sysSize {N_x \times N_y}
\def \droRadius {R_0}

\def \pLap {p_{_\mathrm{Lap}}}
\def \density {\rho}
\def \denLiquid {\rho_l}
\def \denGas {\rho_g}

\def \surTensionLG {\gamma_{lg}}

\def \kinViscosityL {\nu_l}

\def \dynViscosity {\mu}

\newcommand{\vect}[1]{\textbf{#1}}
\newcommand{\Order}[1]{\mathcal{O}{#1}}

\def \Oh {\mathrm{Oh}}

\begin{document}

\markboth{S. Srivastava et al.}
{Lattice boltzmann method to study the contraction of a viscous ligament}

\catchline{}{}{}{}{}

\title{LATTICE BOLTZMANN METHOD TO STUDY THE CONTRACTION OF A VISCOUS LIGAMENT}

\author{{SUDHIR SRIVASTAVA$^{1,\dagger}$}, {THEO DRIESSEN$^2$}, {ROGER JEURISSEN$^{1,3}$}, {HERMAN WIJSHOFF$^4$}, and {FEDERICO TOSCHI$^{1,\ddagger}$}}

\address{$^1$Department of Applied Physics,
  Eindhoven University of Technology, \\P.O. Box 513, 5600 MB Eindhoven, The
  Netherlands\\$^\dagger$s.srivastava@tue.nl, $^{\ddagger}$f.toschi@tue.nl\\
$^2$Faculty of Science and Technology, University of Twente,\\
P.O. Box 217
7500 AE Enschede
The Netherlands\\T.W.Driessen@utwente.nl\\
$^3$ACFD consultancy,
Sint Camillusstraat 26, 6045 ES Roermond, The
  Netherlands\\roger@acfd-consultancy.nl\\$^4$ Oc\'{e} Technologies B.V., P.O. Box 101, 5900 MA, Venlo, The Netherlands\\herman.wijshoff@oce.com}

\maketitle

\begin{history}
\received{Day Month Year}
\revised{Day Month Year}
\end{history}

\begin{abstract}
We employ a recently formulated axisymmetric version of the multiphase
Shan-Chen (SC) lattice Boltzmann method (LBM) [Srivastava \etal, in preparation (2013)] to simulate the contraction of a liquid
ligament. We compare the axisymmetric LBM simulation against the slender jet (SJ) approximation model [T. Driessen and R. Jeurissen, IJCFD {\bf 25}, 333 (2011)]. We compare the retraction dynamics of the tail-end of the liquid ligament from the
LBM simulation, the SJ model, Flow3D simulations and a simple model based on the force balance (FB). We find good agreement between the theoretical prediction (FB),  
the SJ model, and the LBM simulations.
\end{abstract}

\ccode{PACS Nos.: 11.25.Hf, 123.1K}

%
%

\section{Introduction}
\begin{figure}[!h]
\centerline{\psfig{file=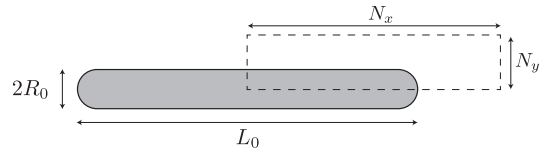,width=7cm}}
\vspace*{8pt}
\caption{A schematic of the initial configuration of the axisymmetric viscous ligament. The rectangular dotted box of size $\sysSize$ represents the domain for LBM simulation.}
\label{fig:filament_schamatics}
\end{figure}
The formation of liquid ligaments is ubiquitous, it happens whenever there is a droplet fragmentation.\cite{Villermaux2007} Examples 
of fragmentation processes are the breakup of a liquid ligament stretched from a bath or the collapse of a liquid film.\cite{Marmottant2004b,Bremond2005} The formation of these liquid ligaments is very common in the breakup of ocean spume where they influence the properties of the marine aerosols.\cite{Veron2012}  In industry, the dynamics of ligaments is a key issue for the print quality in inkjet printing,\cite{Wijshoff2010} where elongated liquid ligaments are ejected from the nozzle (see Fig.~\ref{fig:filament_schamatics}). For optimal print quality, the ligaments should contract to a single droplet before they hit the paper.
Depending on the fluid properties, size and shape of the ligament, it may collapse into a single droplet (stable contraction), or it may break up 
into several droplets (unstable contraction).\cite{Eggers2008} The stability of the contraction of a smooth ligament crucially depends on 
the Ohnesorge number,${\Oh = \kinViscosityL\sqrt{\denLiquid/(\surTensionLG  \droRadius)}}$,\cite{Schulkes2006,Stone1986,Notz2004} where 
$\kinViscosityL$, $\denLiquid$,$\surTensionLG $, and $\droRadius$ are the kinematic viscosity, fluid density, surface tension and radius of the 
ligament, respectively. When ${\Oh > \Order(0.1)}$, the viscous dissipation dominates and there is no energy left to deform the surface 
of the ligament, hence the contraction always remains stable. On the other hand when $\Oh < \Order(0.1)$, low viscous dissipation allows for 
large surface deformation that may result in the breakup of the ligament. Notz \etal  found that the stability of contraction for $\Oh < \Order(0.1)$ depends on the aspect ratio of the ligament, $\Gamma_0= L_0/(2R_0)$, where $L_0$ is the initial length of the ligament, and for ${\Oh 
=  \Order{(0.1)}}$ the contraction of the ligament is stable and independent $\Gamma_0$.\cite{Notz2004} In this work we use the axisymmetric multiphase LBM to simulate the stable contraction of the ligament.\cite{Srivastava2013} We validate the LBM model by 
comparing it against the 1D numerical slender jet (SJ) model by Driessen \& Jeurissen,\cite{Driessen2011} an analytic model based on force 
balance (FB), and the Flow3D\footnote{Flow3D$^{\mathrm{TM}}$ is CFD software developed by Flow Science Inc., Santa Fe, New Mexico.} 
simulation.

%
\subsection{Lattice Boltzmann method}
In this section we prescribe a brief description of the axisymmetric LBM for multiphase flow.\cite{Srivastava2013} The model is defined on the Cartesian two-dimensional (2D) lattice by means of the nine-speeds, $\vect{c}_i \equiv (\cix, \ciy)$, and distribution function, $f_i$:
\begin{equation} \label{eq:lbm2d}
  f_i(\vect{x}+ \vect{c}_i\delta t, t+\delta t) = f_i(\vect{x},t) -  \frac{1}{\tau} \big( f_i(\vect{x},t)-\feq_i(\density,\ueq)
  \big)+ \delta t\,h_i\,,
\end{equation}
where $\vect{x} = (x, y)$ is the position vector, $t$ is time and $\delta t$ is the time step. In the above expression we have made use of the 
BGK approximation to let the distribution relax to the equilibrium distribution, $\feq_i$. The bulk viscosity, $\dynViscosity$, of the fluid is related
to the relaxation parameter, $\tau$, as $\dynViscosity = \density \cs^2 \delta \timStep \left(\tau - 0.5\right)$, where $\cs = \sqrt{3}$ is the speed 
of sound in the LB model.  The fluid density, $\density$ and velocity ${\vect{u} \equiv  (\velX, \velY)}$ are defined as:
\begin{align}
\density&=\sum_{i}f_i\,, & \vect{u} &=\frac{1}{\density}\sum_i{\ci f_i}\,,
\end{align}
respectively. In absence of any external force $\ueq = \vect{u}$. The additional term $h_i$ in Eq.~(\ref{eq:lbm2d}) 
has the following form:
\begin{equation}\label{eq:sourceterm}
h_i = W_i\Big(-\frac{\density \velY}{y} + \frac{1}{yc^2_s}\big(\cix h_{ix} + \ciy h_{iy}\big)\Big),
\end{equation}
where $(h_{ix} ,\,  h_{iy}) = \Big(\cix\big( \dynViscosity \big(\dy \velX + \dx \velY\big)  - \density \velX \velY \big),\, \ciy\big( 2\dynViscosity \big(\dy \velY -\posY^{-1}\velY\big)  -\density \velY^2 \big)\Big)$ and $W_i$'s are the lattice dependent weights.
The Chapmann-Enskog expansion of Eq.~(\ref{eq:lbm2d}) gives the axisymmetric continuity and Navier-Stokes' equations (NS):
 \begin{align}\label{eq:axis_cty}
       \dt \rho +\nabla\cdot(\rho \vect{u})& = -y^{-1}\rho \velY, 
\end{align}
and
\begin{align}\label{eq:axis_ns}
	\dt \big(\rho\vect{u}\big) + \nabla\cdot\big(\rho\vect{u}\vect{u}\big) = -\nabla p + \nabla\cdot\big(\mu\big(\nabla\vect{u} + \nabla\vect{u}^{T}\big)\big) + \vect{f}\,,\,\,\,\,\,\\
\text{where\,\,\,\,}\vect{f} = \posY^{-1}\Big( \dynViscosity \big(\dy \velX + \dx \velY\big)  - \density \velX \velY\,,\, 2\dynViscosity \big(\dy \velY -\posY^{-1}\velY\big)  -\density \velY^2 \Big),\nonumber
\end{align}
and $\nabla$ is the 2D divergence operator in the Cartesian coordinate system.\cite{Srivastava2013,Landau1984} In this manuscript, symbols $x$ and $y$ represent the axial and radial distances, respectively. The Eqs.~(\ref{eq:axis_cty}) and (\ref{eq:axis_ns}) are written in 
a form to emphasize the 2D continuity and NS equation. The additional term ${-y^{-1}\rho \velY}$ and $\vect{f}$ on R.H.S. of Eq.~(\ref{eq:axis_cty}) and (\ref{eq:axis_ns}), respectively, arise due to axisymmetry. 

The long-range interaction force, $\vect{F}$, in the Shan-Chen (SC) model is defined as:
\begin{equation}\label{eq:scforce}
 \vect{F} = -G c_s^2\,\delta t\,\psi \hat{\nabla} \psi - \frac{G}{2} c_s^4\,(\delta t)^3\,\psi \hat{\nabla}(\hat{\nabla}^2 \psi) + \Order((\delta t)^5),
\end{equation}
where $G$ is the interaction strength between two phases and $\hat{\nabla}, \hat{\nabla}^2$ are the gradient and Laplace operators, respectively in the 3D Cartesian coordinate system.\cite{Shan1993,Shan1994} The Eq.~(\ref{eq:scforce}) for axisymmetric cylindrical polar coordinate is given by:
\begin{equation}\label{eq:scforce2}
 \vect{F} = -\scG c_s^2\,\delta t\,\psi \nabla \psi - \frac{\scG}{2} c_s^4\,(\delta t)^3\,\psi \nabla\Big(\nabla^2 \psi + \posY^{-1} \dy \psi\Big) + \Order((\delta t)^5),
\end{equation}
where $\nabla^2$ is the 2D Laplace operator in the Cartesian coordinate system. The axisymmetric 
contribution in addition to the 2D SC force comes from second term of the Eq.~(\ref{eq:scforce}) and it is given by ${- \frac{\scG}{2} c_s^4(\delta t)^3\psi \nabla (\posY^{-1} \dy \psi)}$.

The force \vect{F} given by Eq.~(\ref{eq:scforce2}) is added in to the system by shifting the equilibrium velocity as ${\ueq =\frac{1}
 {\density} \left(\sum_i {\bm c}_if_i + \tau\vect{F}\right)}$, and the fluid velocity is defined as ${\vect{u}=\frac{1}{\density} \left(\sum_i \vect{c}_if_i 
 + \frac{\delta t}{2} \vect{F}\right)}$. The finite difference approximations used for the derivatives in Eqs.~(\ref{eq:sourceterm}), (\ref{eq:scforce2}) are isotropic and fifth-order accurate. This is necessary in order to minimize the truncation error that appears in the long-
 wavelength and in the  small Mach number limit of Eq.~(\ref{eq:lbm2d}). The non-ideal pressure, ${p_{NI} = c_s^2 \density + \frac{c_s^2\scG}
 {2}\psi^2}$ in the axisymmetric multiphase LBM is same as the non-ideal pressure for 3D LBM.\cite{He2002} Our choice of the effective 
 density functional is ${\psi( \density) = \density_0\big(1 -\exp(-\density/\density_0)\big)}$, where $\density_0$ is a reference density.
 
\subsection{Lubrication Theory model}
We are using the slender jet approximation to model the stability of an axisymmetric viscous liquid ligament.\cite{Notz2004,Driessen2011,Eggers1994,Hoeve2010,Shi1994,Wilkes1999,Garcia2008} In the slender jet approximation, the fluid
flow in the axial direction is assumed to be dominant. Therefore, radial inertia is neglected and the axial velocity is assumed to be uniform in the radial direction. As a result, the fluid interface is a well defined, single valued function of the axial coordinate, from which the full curvature  of the interface can be calculated. If we use the initial radius of ligament, $R_0$, as the length scale and the capillary time, $t_{cap} = \sqrt{\density_lR_0^3/\gamma_{lg}} $ as the time scale, then the SJ model in the dimensionless form is given by:
\begin{align}\label{eq:SJmodel}
 \dt \posH &= - \velX \dx \posH - \frac{1}{2}\posH \dx \velX, \,\,\,\,\,
 \dt \velX = -\velX \dx \velX - \dx \pLap + 3\, \Oh\, \posH^{-2}\dx(\posH^2 \dx \velX ), \\
\pLap &= \posH^{-1}\Big(1+(\dx \posH)^2\Big)^{-1/2} -{\dxx}\posH{\Big(1+(\dx \posH)^2\Big)^{-3/2}}, \nonumber 
\end{align}
where $\posH, \velX, \posX, \timStep$ and $\pLap$ are dimensionless, and represent the radius of the jet, axial velocity, axial coordinate, time, 
and Laplace pressure, respectively. For this study we use the numerical model developed by Driessen and Jeurissen to solve Eq.\,(\ref{eq:SJmodel}).\cite{Driessen2011} The solutions to these equations are singular at each pinch-off, and at each collision of liquid bodies.\cite{Eggers1994} To allow the described physical system to transfer across the singularities that occur at pinch-off and coalescence, the surface tension term is regularized by a modification at a radius of the order of the cutoff radius, $h_c$. 
The cutoff radius, $h_c$ is a control parameter of the regularization, and is chosen to scale with the spatial step. For the SJ simulations
presented in this manuscript $h_c= \droRadius/60$.

\section{Results and discussion}
In this section we show the comparison of simulation from LBM and  SJ for the contraction of liquid ligament. The LBM simulation is carried out
for the following parameters (LBM units): system size, $\sysSize = 1600 \times 256$, $L_0 = 2000$, $\droRadius = 49.5$, relaxation
parameter, $\tau = 1$, kinematic viscosity, $\kinViscosityL = 0.17$, Shan-Chen interaction parameter, $\scG = -5$, liquid density, $\denLiquid = 1.95$,
vapor density, $\denGas = 0.16$, and surface tension, $\surTensionLG= 0.0568$. For above LBM parameters we have $\Oh = 0.14$, $
\Gamma_0 = 20$. This parameter choice is suitable for simulating the stable contraction of a smooth ligament. For our study it is sufficient to simulate only half of the liquid ligament (see Fig.~\ref{fig:filament_schamatics}). We use the symmetry boundary condition at left, right and bottom boundaries and the free slip at the top boundary.\cite{Succi2001} 

In order to make a comparison between the two models we need to have a common system for measuring the physical quantities and we 
opted for expressing quantities in dimensionless units. We choose the initial radius of the ligament, $R_0$, and the capillary time, $t_{cap}$, to 
scale length and time in LBM simulations. For SJ simulations we use the aspect ratio, $\Gamma_0 = 20$, and the $\Oh =
0.14$.

\begin{figure}[!h] 
\centerline{\psfig{file=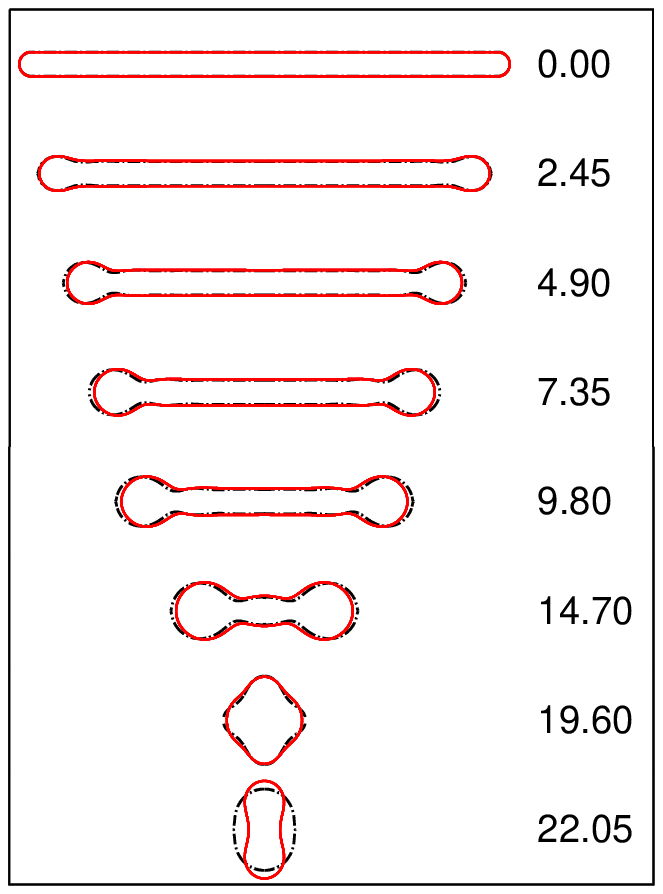,width=0.55\textwidth}\label{fig:compare_interface}
{\psfig{file=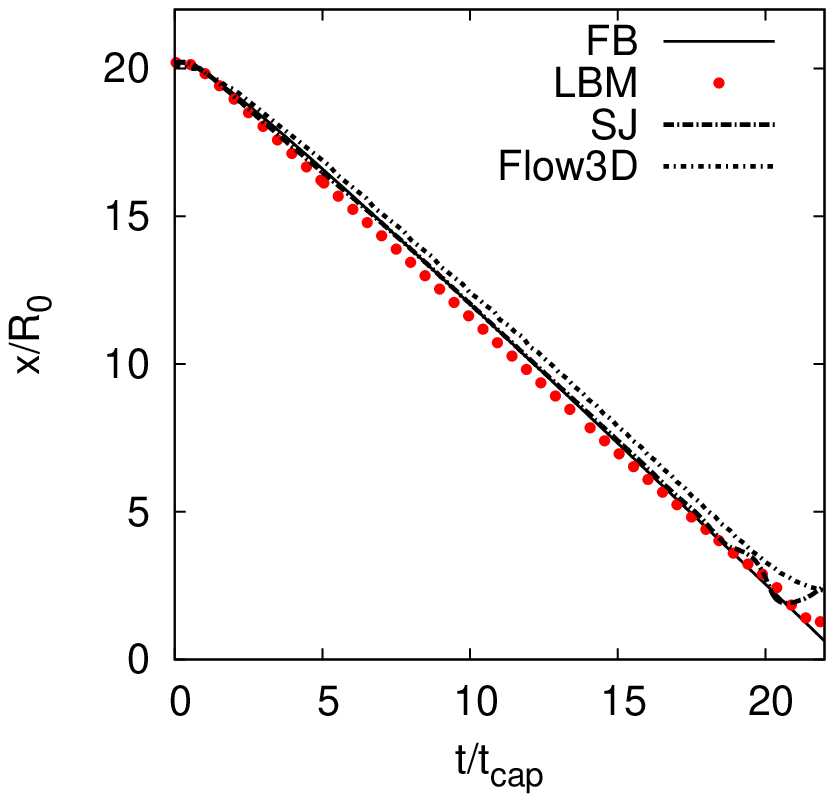,width=0.55\textwidth}}}\label{fig:tail_velocity}
\vspace*{8pt}
\caption{Left panel: Time evolution of interface profile of the liquid ligament. The labels on the figure show the dimensionless time,  $t/t_{cap}$. 
The data points from the LBM simulation are shown in red color, whereas the data from SJ model are shown in black color. Right panel: The tip 
location of the collapsing filament as a function of time in the presented models. The difference between the LBM simulation, SJ simulation 
and FB model is smaller than the interface thickness in the LBM simulation. The simulations and the analytical result agree with each other, up 
to the moment when the tail droplets merge.}
\end{figure}

First, we compare the time evolution of the ligament shape obtained from the LBM and the SJ simulation (see Fig.~\ref{fig:compare_interface}). 
During the collapse, there is a perfect agreement of all the models. When the tail droplets merge into one big droplet, the simulation 
results start to differ; in the LB simulation, the maximum radial extent of the droplet is larger and dimples form on both sides of the droplet. We hypothesize that this is due to the lubrication approximation in the SJ model. When the tail droplets merge, ${\dy\velX}$ becomes significant, while it is neglected in the SJ model. When the radial extent of the droplet reaches its maximum, the kinetic energy is mostly converted into surface energy. A smaller radial extent indicates that the dissipation was larger. The origin of this numerical dissipation is similar to the dissipation in a shock in gas dynamics, or a hydraulic jump in hydraulic engineering; momentum is conserved, but energy is dissipated in a shock. 
The concave drop shape obtained in the LB simulation indicates that the lubrication approximation causes dissipation here. This shape cannot be represented as a single valued function in the one dimensional space of the SJ model, and the numerical dissipation in the SJ model is the effect that prevents the formation of these dimples.
For the second validation we compare our LBM results to the SJ model and the Flow3D simulation. Additionally, we estimate the position of
the tail-end of the ligament by an analytical model based on the force balance  (FB).

In the FB model the rate of change of the mass, $m$, and momentum, $P = m\velX$, of the tail-drop is given by:
 \begin{align}\label{eq:ode}
 \frac{dx}{dt} &= \velX, & \frac{dm}{dt} &= \denLiquid \pi R^2\velX , & \frac{dP}{dt} &=  -\pi R^2\frac{\surTensionLG}{R} = -\pi \surTensionLG R  
 \end{align}
where $\velX$ is the velocity of the tail-drop, $2x$ is the length and $R$ is the radius of the ligament. The solution of Eq.~(\ref{eq:ode}) subject 
to the initial conditions: ${x(0) = 0.5L_0 -R_0}$, ${ m(0) =  (2\pi/3) \denLiquid R_0^3}$ and ${P(0) =  0}$,  gives us the length of the ligament in time,
$2x(t)$ ${(R_0 = R(0))}$. In this force balance the tail velocity converges to the capillary velocity,  $u_{cap}=\sqrt{\gamma_{lg}/(\rho_l R)}$.\cite{Schneider1966} The solutions from FB model, SJ model, Flow3D simulation and LBM simulation are in very good agreement with 
each other (See Fig. \ref{fig:compare_interface}, right panel).

\section{Conclusion}
The axisymmetric multiphase SC LBM has been validated on the test problem of the stable contraction of liquid ligament.\cite{Srivastava2013}  
For this validation the LBM simulations was compared to SJ, FB models, and Flow3D simulations. Furthermore the position of the tail-end of 
the drop was compared with a model based on the balance of forces. We found that the proposed axisymmetric multiphase SC LBM can 
accurately simulate the collapse of viscous liquid ligament.\cite{Srivastava2013} 
\section*{Acknowledgments}
This work is part of the research program of the Foundation for Fundamental Research on Matter (FOM), which is part of the Netherlands 
Organization for Scientific Research (NWO).



\begin{thebibliography}{00}
\bibitem{Villermaux2007}
E. Villermaux, {\it Ann. Rev. Fluid Mech.} {\bf {39}},  419  (2007).

\bibitem{Marmottant2004b}
P. Marmottant and E. Villermaux, {\it Phys. Fluids} {\bf 16},  2732  (2004).

\bibitem{Bremond2005}
N. Bremond and E. Villermaux, {\it J. Fluid Mech.} {\bf 524},  121  (2005).

\bibitem{Veron2012}
Veron, F. C. \etal, {\it Geophys. Res. Lett.} {\bf  39},  L16602  (2012).

\bibitem{Wijshoff2010}
H. Wijshoff, {\it Phys. Rep.} {\bf 491},  77   (2010).

\bibitem{Eggers2008}
J. Eggers and E. Villermaux, {\it Rep. Prog. Phys.} {\bf 71},  036601  (2008).

\bibitem{Schulkes2006}
R. M. S. M. Schulkes, {\it J. Fluid Mech.} {\bf 309}, 277 (2006).

\bibitem{Stone1986}
H. Stone \etal, {\it J. Fluid Mech.} {\bf 173}, 131Ð158 (1986).

\bibitem{Notz2004}
P.~K. Notz and O.~A. Basaran, {\it J. Fluid Mech.} {\bf 512},  223  (2004).

\bibitem{Srivastava2013}
S. Srivastava {\etal}, in preparation (2013).

\bibitem{Driessen2011}
T. Driessen and R. Jeurissen, {\it Int. J. Comput. Fluid Dyn.} {\bf 25},  333  (2011).

\bibitem{Shan1993}
X. Shan and H. Chen, {\it Phys. Rev. E} {\bf 47},  1815  (1993).

\bibitem{Shan1994}
X. Shan and H. Chen, {\it Phys. Rev. E} {\bf 49},  2941  (1994).

\bibitem{Landau1984}
L. D. Landau and E. M. Lifshitz, {\it Fluid Mechanics} {\bf 6},(Pergamon,1959)

\bibitem{He2002}
X. He and G. Doolen, {\it J. stat. phys.} {\bf 107},  309  (2002).  

\bibitem{Eggers1994}
J. Eggers and T.~F. Dupont, {\it J. Fluid Mech.} {\bf 262},  205  (1994).

\bibitem{Hoeve2010}
W. van Hoeve {\it et~al.}, {\it Phys. Fluids} {\bf 22}, Ê122003 Ê(2010).

\bibitem{Shi1994}
X.~D. Shi \etal, Science {\bf 265},  219   (1994).

\bibitem{Wilkes1999}
E.~D. Wilkes \etal, {\it Phys. Fluids} {\bf 11},  3577  (1999).

\bibitem{Garcia2008}
F.~J.~Garc\'{i}a, and H.~Gonz\'{a}lez, {\it J. Fluid Mech.} {\bf 602}, 81 (2008).
  
\bibitem{Succi2001}
S.~Succi, {\it The Lattice Boltzmann Equation for Fluid Dynamics and Beyond}, (Oxford University Press: 2001)

\bibitem{Schneider1966}
J.~M. Schneider, N.~R. Lindblad, C.~D. Hendricks, and J.~M. Crowley, {\it J. Appl.
  Phys. } {\bf 38},  2599  (1966).



\end{thebibliography}
\end{document}